# Remarks on Our Understanding of Quantum Mechanics

Elias P. Gyftopoulos
*Massachusetts Institute of Technology*
*77 Massachusetts Avenue; Room 24-111*
*Cambridge, MA 02139 USA*
(Dated: March 19, 2002)

Interpretations of key concepts, such as uncertainty relations, kinetic energy, value of an observable, probability distributions, the projection or collapse of a wave function postulate, and discrete versus continuous values, that appear in several excellent textbooks on quantum mechanics are reviewed and found not to be consistent with our current understanding of quantum theory. Possible alternatives are suggested.

I.  INTRODUCTION

In the Appendix of this paper, we provide a brief summary of definitions, postulates, and major theorems of quantum mechanics. They are based on slightly modified statements made by Park and Margenau [1]. The modifications are primarily due to the recognition by Hatsopoulos and Gyftopoulos [2] that there exist two classes of quantum problems. In the first class, the quantum-mechanical probabilities associated with measurement results are fully described by wave functions or projectors, whereas in the second class the probabilities just cited require density operators that involve no statistical averaging over projectors – no mixture of quantum and statistical probabilities. The same result emerges from the excellent review of the foundations of quantum mechanics by Jauch [3]. Implications and applications of purely quantum-mechanical density operators are discussed in several publications [4-7].

In light of the statements given in the Appendix, it behooves us to reexamine the interpretations prevailing in the scientific literature about the concepts of uncertainty relations, kinetic energy, value of an observable, probability distributions, the projection or collapse of a wave function postulate, and discrete versus continuous values and classical mechanics, and see whether they are consistent with the theory. We show that they are not.

II.  UNCERTAINTY RELATIONS AND ACCURACY OF MEASUREMENT RESULTS

Ever since the inception of quantum mechanics, the uncertainty relation that corresponds to a pair of observables represented by noncommuting operators is interpreted as a limitation on the accuracy with which the observables can be measured. For example, Heisenberg [8] says: "… processes of atomic physics can be visualized equally well in terms of waves or particles". Thus the statement that the position of an electron is known to within a certain accuracy $\Delta x$ at the time $t$ can be visualized by the picture of a wave packet in the proper position with an approximate extension $\Delta x$. By "wave packet" is meant a wavelike disturbance whose amplitude is appreciably different from zero only in a bounded region. This region is, in general, in motion, and also changes its size and shape, i.e., the disturbance spreads. The velocity of the electron corresponds to that of the wave packet, but this latter cannot be exactly defined, because of the diffusion which takes place. This indeterminateness is to be considered as an essential characteristic of the electron, and not as evidence of the inapplicability of the wave picture. Defining momentum as $p_x = \mu v_x$ (where $\mu$ = mass of electron, $v_x = x-$component of velocity), this uncertainty in the velocity causes an uncertainty in $p_x$ of amount $\Delta p_x$; from the simplest laws of optics, together with empirically established law $\lambda = h/p$, it can be readily shown that

$$\Delta x \Delta p_x \geq h \quad (1)$$

"This uncertainty relation specifies the limits within which the particle picture can be applied. Any use of the words "position" and "velocity" with an accuracy exceeding that given by equation (1) is just as meaningless as the use of words whose sense is not defined."

"The uncertainty relations can also be deduced without explicit use of the wave picture, for they are readily obtained from the mathematical scheme of quantum theory …"

Again, Louisell [9] says: "… The failure of classical mechanics to account for many experimental results led physicists to the realization that classical concepts were inherently inadequate to describe the physical behavior of events on an atomic scale. To explain these phenomena, a fundamental departure from classical mechanics was necessary. This departure took the form of postulating, as a fundamental law of nature, that there is a limit to the accuracy with which a measurement (or observation) on a physical system can be made. That is, the actual measurement itself disturbs the system being measured in an uncontrollable way, regardless of the care, skill, or ingenuity of the experimenter. The disturbance produced by the measurement in turn requires modification of the classical concept of causality, since, in the classical sense, there is a causal connection between the system and the measurement. This leads to a theory in which one can predict only the probability of obtaining a certain result when a measurement is made on a system rather than an exact value, as in the classical case. … In quantum mechanics, the precise measurement of both coordinates and momenta is not possible even in principle because of the disturbance caused by a measurement."

The remarks by Heisenberg and Louisell are representative of the interpretations of uncertainty relations discussed in practically every textbook on quantum mechanics [10]. Even if they are correct, these interpretations are not warranted and cannot be deduced from uncertainty relations. The reasons for these conclusions are the measurement result and the probability theorems discussed in the Appendix.

The measurement result theorem avers that the measurement of an observable is a precise (perturbation free) and, in many cases, precisely calculable eigenvalue of the operator that represents the observable. So neither a measurement perturbation nor a measurement error is contemplated by the theorem. An outstanding example of measurement accuracy is the Lamb shift [11].

The probability theorem avers that we cannot predict which precise eigenvalue each measurement will yield except in terms of either a prespecified or a measurable probability or frequency of occurrence.

It follows that an ensemble of measurements of an observable performed on an ensemble of identical systems, identically prepared yields a range of eigenvalues, and a probability or frequency of occurrence distribution over the eigenvalues. In principle both results are precise and involve no disturbances induced by the measuring procedures.

To be sure, each probability distribution of an observable represented by operator X has a variance

$$(\Delta X)^2 = \text{Tr}\rho X^2 - (\text{Tr}\rho X)^2 = \langle X^2 \rangle - \langle X \rangle^2 \tag{2}$$

and a standard deviation $\Delta X$, where $\rho$ is the projector or density operator that describes all the probability distributions of the problem in question. Moreover, for two observables represented by two noncommuting operators A and B, that is,

$$AB - BA = iC \tag{3}$$

it is readily shown that $\Delta A$ and $\Delta B$ satisfy the uncertainty relation

$$\Delta A \Delta B \geq |\langle C \rangle|/2 \tag{4}$$

It is evident that each uncertainty relation refers neither to any errors introduced by the measuring instruments nor to any particular value of a measurement result. The reason for the latter remark is that the value of an observable is determined by the expectation value of the operator representing the observable and not by any individual measurement result (see Section IV, and the definition of state in the Appendix).

III. KINETIC ENERGY

The algebraic expression for kinetic energy is not the same for all paradigms of physics. For example, in classical mechanics, the kinetic energy of a particle of mass $m_0$ in a force field derived from a potential function $V(r)$ is shown to be

$$(\text{K.E.})_{cl} = m_0 v^2 / 2 \tag{5}$$

where $v$ is the speed of the particle. Again, in special relativity

$$(\text{K.E.})_{sr} = E - m_0 c^2 \neq m_0 v^2 / 2 \tag{6}$$

where $E$ is the energy of the particle, and $c$ the speed of light. Again, in general relativity only energy is defined, and no distinction can be made between different kinds of energy.

The radical differences in three outstanding paradigms of physics just cited beg the question: "What is the analytical expression for kinetic energy in quantum mechanics?"

Practically all publications on quantum mechanics suggest that the answer is $\langle p^2 \rangle / 2m_0$, and provide as evidence the comparison of the virial theorem of classical mechanics with results obtained for the hydrogen atom [12].

In what follows, we discuss seven reasons which show that such an answer is completely unjustified, and leads to monstrous contradictions. (i) In all paradigms of physics, motion ($\kappa\iota\nu\eta\sigma\mathrm{is}$ = kinesis) is defined in terms of a nonzero value of velocity, and velocity is defined in terms of the spatial and time coordinates which are common to all paradigms. So, it is a contradiction to have a particle with a quantum-mechanical value of velocity $\langle p \rangle / m_0 = 0$, and yet assign to such a particle a nonzero kinetic energy $\langle p^2 \rangle / 2m_0 \neq 0$; (ii) Three energy eigenfunctions of a free particle moving in the *x*-direction are

$$\exp(ip_1 x) + \exp(-ip_1 x): \quad \langle p \rangle = 0; \quad E_1 = p_1^2 / 2m_0 \tag{7}$$

$$\exp(ip_1 x): \quad \langle p \rangle = p_1; \quad E = \langle p^2 \rangle / 2m_0 = p_1^2 / 2m_0 \tag{8}$$

$$\exp(-ip_1 x): \quad \langle p \rangle = -p_1; \quad E = \langle p^2 \rangle / 2m_0 = p_1^2 / 2m_0 \tag{9}$$

Despite the fact that all three energy eigenfunctions correspond to the same energy, we know that only the first is a *stationary energy eigenfunction* which involves no motion, whereas the second and third are *steady-state energy eigenfunctions*, that is, the particle is moving with momentum either $p_1$ or $-p_1$, respectively; (iii) Prior to inducing a current in a superconductor at temperature *T*, no magnetic field is observed. Upon inducing a current (motion of electrons), a magnetic field is created. Even in the absence of current $\langle p^2 \rangle \neq 0$. So, if $\langle p^2 \rangle$ is an indicator of motion, some magnetic field ought to have been observed even in the absence of electron motion, that is, even if $\langle p \rangle = 0$; (iv) The nth energy eigenfunction of a particle in an infinitely deep potential well extending over $-a \leq x \leq a$, where n = 1, 2, ..., has $\langle p \rangle = 0$, $\langle p^2 \rangle \neq 0$, and energy eigenvalue $\varepsilon_n = \langle p^2 \rangle / 2m_0$. Because the potential within a well is zero, if we interpret $\langle p^2 \rangle / 2m_0$ as kinetic energy, and use the classical mechanics result that energy is the sum of kinetic and potential energies, we end up with the following monstrosity. Upon making energy measurements on an ensemble of identical systems each of which is characterized by the nth energy eigenfunction, each measurement yields the same eigenvalue $\varepsilon_n$, and therefore the probability density function is $\delta(\varepsilon - \varepsilon_n)$, where $\delta$ is the Dirac delta function. On the other hand, upon making an ensemble of momentum measurements, and calculating the expectation value of $p^2 / 2m_0$, we find $\langle p^2 \rangle / 2m = \varepsilon_n$, but the probability density function of both *p* and $p^2$ is $|\phi_n(p)|^2$ and not $\delta(\varepsilon - \varepsilon_n)$, where $\phi_n(p)$ is the Fourier transform of the energy eigenfunction corresponding to $\varepsilon_n$. The Fourier transform extends from $-\infty$ to $+\infty$. Thus, we end up with the monstrosity that the same observable, energy, has simultaneously two probability density functions, $\delta(\varepsilon - \varepsilon_n)$ and $|\phi_n(p)|^2$!; (v) In a change of state from $A_1$ to $A_2$, momentum conservation requires that the momentum transfer be $\langle p_2 \rangle - \langle p_1 \rangle$ and not $\langle p_2^2 \rangle^{1/2} - \langle p_1^2 \rangle^{1/2}$. Similarly, energy conservation requires that the energy transfer be $\langle H_2 \rangle - \langle H_1 \rangle$ and not $\langle H_2^2 \rangle^{1/2} - \langle H_1^2 \rangle^{1/2}$; (vi) In general, in order to use a classical expression we must verify that we can replace an expectation value by a classical functional relation, that is $\langle A(x, p) \rangle = a(x_0, p_0)$, where A is the operator corresponding to an observable, $a$ the classical function that corresponds to operator A, $x_0 = \langle x \rangle$, and $p_0 = \langle p \rangle$ (Shankar [13]). For the case $A = p^2$, we cannot write $\langle p^2 \rangle = \langle p_0^2 \rangle = \langle p \rangle^2$ because $\langle p^2 \rangle - \langle p \rangle^2 = (\Delta p)^2$. The variance $(\Delta p)^2$ becomes zero if and only if a free particle is characterized by a momentum – steady state – eigenfunction. So, in general we cannot use the classical algebraic expression to represent kinetic energy in quantum mechanics; (vii) According to the definition of state (see Appendix), an indicator of motion is the value of momentum $\langle p \rangle$. Specifically, if $\langle p \rangle = 0$ the particle is not moving because then classically $p_0 = 0$. If $\langle p \rangle \neq 0$ the particle is moving because then classically $p_0 \neq 0$. Said differently, $\langle p^2 \rangle$ is always positive and not an indicator of motion. It is noteworthy that $\langle p \rangle = 0$ is not an average over many particles moving in different directions. It is the value of the momentum of one particle.

In view of the preceding considerations, we must conclude that in quantum mechanics, an algebraic expression for kinetic energy has not been, and perhaps cannot be defined in the same sense that, in general relativity, kinetic energy is not and cannot be defined.

IV.  EXPECTATION VALUE AND VALUE OF AN OBSERVABLE

More often than not the concept of expectation value of an observable is introduced as a tool that has no special theoretical and experimental significance, but simply avoids difficult or almost impossible calculations. For example, Shankar [14] says: "Given a large ensemble of N particles in a state $|\psi\rangle$, quantum theory allows us to predict what fraction will yield a value $\omega$ if the variable $\Omega$ is measured. This prediction, however, involves solving the eigenvalue problem of the operator $\Omega$. If one is not interested in such detailed information on the state, one can calculate instead an average over the ensemble called the expectation value $\langle \Omega \rangle$".

Such characterization of expectation values overlooks and annuls both their theoretical and their experimental importance. The definition of state (see Appendix) indicates that a complete set of linearly independent expectation values is uniquely equivalent to the probabilities or frequencies of occurrence associated with all measurement results and, conversely, that probabilities or frequencies of occurrence associated with all measurement results are uniquely equivalent with a complete set of linearly independent expectation values. Said differently, the mappings from a density operator $\rho$ to expectation values, and from expectation values to a density operator $\rho$ are linear. Moreover, and perhaps more importantly, because expectation values in principle can be measured, they provide the means for the determination of the probabilities associated with the measurement results.

In addition, expectation values are exclusively important in each conservation principle and associated balance equation, such as the principle of energy conservation and the associated energy balance, and the principle of momentum conservation, and the associated momentum balance.

In view of these observations, we must conclude that each expectation value is the value of the observable at the instant in time at which the measurements are made.

## V. PROJECTORS AND DENSITY OPERATORS

Every textbook on quantum mechanics avers that the probabilities associated with measurement results of a system in a state "i" are described by a wave function $\psi_i$ [15] or, equivalently, a projector $|\psi_i\rangle\langle\psi_i| = \rho_i = \rho_i^2$, and that the density operators $\rho > \rho^2$ are statistical averages of projectors, that is, each $\rho$ represents a mixture of quantum mechanical probabilities determined by projectors, and nonquantum-mechanical or statistical probabilities that reflect our inability to make difficult calculations, our lack of interest in details, and our lack of knowledge of initial conditions. Mixtures have been introduced by von Neumann [16] for the purpose of explaining thermodynamic equilibrium phenomena in terms of statistical quantum mechanics (see also Jaynes [17] and Katz [18]).

Pictorially, we can visualize a projector by an ensemble of identical systems, identically prepared. Each member of such an ensemble is characterized by the same projector $\rho_i$, and von Neumann calls the ensemble *homogeneous*. Similarly, we can visualize a density operator $\rho$ consisting of a statistical mixture of two projectors $\rho_1$ and $\rho_2$ by the ensemble shown in Figure 1. In this ensemble, $\rho = \alpha_1\rho_1 + \alpha_2\rho_2$, $\alpha_1 + \alpha_2 = 1$, $\rho \neq \rho_1 \neq \rho_2$, $\rho_1$ and $\rho_2$ represent quantum-mechanical probabilities, $\alpha_1$ and $\alpha_2$ statistical probabilities, and the ensemble is called *heterogeneous* or *ambiguous* [2].

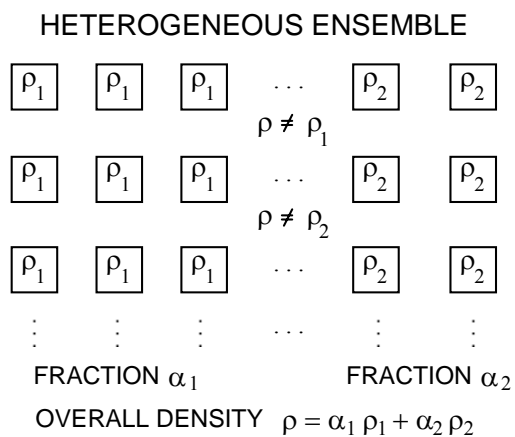

FIG. 1: Pictorial representation of a heterogeneous ensemble. Each of the subensembles for $\rho_1$ and $\rho_2$ can represent either a projector ($\rho_i = \rho_i^2$) or a density operator ($\rho_i > \rho_i^2$).

These results beg the questions: (*i*) are there quantum-mechanical problems that involve probability distributions which cannot be described by a projector but require a purely quantum-mechanical density operator – a density operator which is not a statistical mixture of projectors?; and (*ii*) are such purely quantum-mechanical density operators consistent with the foundations of quantum mechanics?

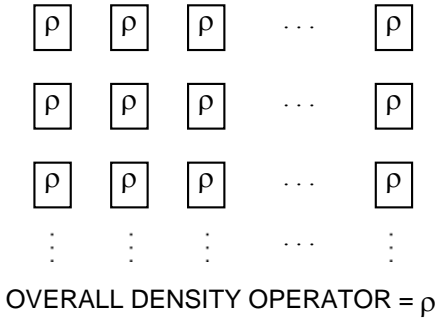

FIG. 2: Pictorial representation of a homogeneous ensemble. Each of the members of the ensemble is characterized by the same density operator $\rho \geq \rho^2$. It is clear that any conceivable subensemble of a homogeneous ensemble is characterized by the same $\rho$ as the ensemble.

Upon close review of the definitions, postulates and key theorems of quantum theory, we find that the answers to both questions are yes (see Appendix). These answers were discovered by Hatsopoulos and Gyftopoulos [2] in the course of their development of a unified quantum theory of mechanics and thermodynamics, and by Jauch [3] in his systematic and rigorous analyses of the foundations of quantum mechanics.

Pictorially, we can visualize a purely quantum-mechanical density operator $\rho > \rho^2$ by an ensemble of identical systems, identically prepared. Each member of such an ensemble is characterized by the same $\rho$ as shown in Figure 2, and by analogy to the results for a projector we call this ensemble *homogeneous* or *unambiguous* [2]. If the density operator is a projector $\rho_i = \rho_i^2$, then each member of the ensemble is characterized by the same $\rho_i$ as proposed by von Neumann.

The recognition of the existence of density operators that correspond to homogeneous ensembles has many interesting implications. It extends quantum ideas to thermodynamics, and thermodynamic principles to quantum phenomena. For example, it is shown that entropy is a measure of the quantum-mechanical spatial shapes of the constituents of a system, and that irreversibility is solely due to the changes of these shapes as the constituents try to conform to the external and internal force fields of the system [5]. Again, it is shown that thermodynamics applies to all systems (both large and small, including one particle systems such as one spin), and to all states (both thermodynamic equilibrium, and not thermodynamic equilibrium), and that entropy is a property of each constituent of a system [4] (in the same sense that inertial mass is a property of each constituent), and not a measure of either ignorance, or lack of information, or disorder [6]. The few results just cited indicate that the restriction of quantum mechanics to problems that require probability distributions described only by projectors is both unwarranted and nonproductive.

It is noteworthy that numerically, say by looking at the matrix elements of a density operator, we cannot decide whether ρ is ambiguous – consists of a mixture of quantum-mechanical and statistical probabilities – or unambiguous – consists of quantum-mechanical probabilities only. The reason is that, in general, a given ρ can be written as a linear combination of different ρ's in an infinite number of ways. In principle, the decision can be made only by subdividing the ensemble representing ρ into subensembles in any conceivable manner, and then examining whether each subensemble is described by the same ρ as the ensemble. Other criteria that can be used to decide whether a ρ is unambiguous – can be represented by a homogeneous ensemble – are given by Hatsopoulos and Gyftopoulos [2].

VI. THE PROJECTION OR COLLAPSE OF THE WAVE FUNCTION POSTULATE

Among the postulates of quantum mechanics, many authoritative textbooks include von Neumann's projection or collapse of the wave function postulate [19, 20].

An excellent, rigorous, and complete discussion of the projection postulate is given by Park [21]. He finds that the postulate is "absurd, false, and useless".

The only thing that we wish to add here is an argument against the postulate based on a violation of the position-momentum uncertainty relation. We consider a structureless particle confined in a one-dimensional, infinitely deep potential well of width *L*. Initially, the particle is in a state characterized by a projector $|\psi\rangle\langle\psi|$. According to the projection postulate, upon a momentum measurement the particle must collapse into a momentum eigenstate. Suppose that the eigenstate just cited is characterized by the ith momentum projector $|p_i\rangle\langle p_i|$. For such a projector, the standard deviation of position measurement results satisfies the relations $0 < \Delta x < L$, and the standard deviation of momentum measurement results $\Delta p = 0$. Accordingly, $\Delta x \Delta p = 0 < \hbar/2$ instead of $\Delta x \Delta p \geq \hbar/2$. In view of the unquestionable validity of uncertainty relations, we must conclude that the projection postulate cannot be a valid postulate of quantum theory.

VII. DISCRETE VERSUS CONTINUOUS VALUES

It is often argued that the transition from quantum mechanics to classical mechanics occurs at large values of energy because then discrete (quantized) values of observables become continuous [22, 23].

Whereas it is true that, for large (nonrelativistic) values of energy, quantum effects are nonmeasurable and unimportant, and classical mechanics describes very accurately our perceptions and observations of physical phenomena, the reason for these facts is not a switch from discrete to continuous values but the narrowing of the probability distribution functions over spatial and momentum measurement results. This narrowing allows the replacement of density operators, $\rho \geq \rho^2$, in Hilbert space by the classical distribution function $\delta(x-x_0)\delta(p-p_0)$ in phase space and, therefore, the transition from quantum to classical mechanics.

If we justify the transition as the change from discrete to continuous values of observables, we misrepresent the three most important characteristics of quantum theory which are (see Appendix): (*i*) upon an ensemble of measurements of an observable on an ensemble of identical systems, identically prepared we obtain a spectrum of measurement results; this spectrum may be discrete, continuous, or a mixture consisting of both discrete and continuous results. For example, a particle in an infinitely deep potential well has a discrete spectrum of energy eigenvalues. Again, the spectrum of position eigenvalues is always continuous; (*ii*) we cannot predict which value each measurement will yield except in terms of a probability or frequency of occurrence; and (*iii*) the value of any observable is determined by an expectation value and, therefore, is always continuous and never discrete.

These observations show very clearly that the transition from quantum to classical mechanics cannot be based on discrete versus continuous values but on a transition from probabilities described by a density operator $\rho$ in Hilbert space to a probability density function $\delta(x-x_0)\delta(p-p_0)$ in phase space. Of course the probability density function has a standard deviation equal to zero and, therefore, the instantaneous value of any observable represented by a Hermitian operator $A(x, p)$ equals the classical function $a(x, p)$.

VIII. CONCLUDING REMARKS

The traditional interpretations of the concepts of uncertainty relations, kinetic energy, value of an observable, density operators, the projection postulate, and discrete versus continuous values of observables are reviewed and found to be inconsistent with our current understanding of quantum theory, and unsuitable for the resolution of the century and a half old question of the relation between mechanics and thermodynamics. Alternative interpretations are suggested.

The most important aspect of this review is the brief discussion of density operators $\rho > \rho^2$ which consist exclusively of quantum-mechanical probabilities and not a mixture of quantum-mechanical probabilities represented by projectors $\rho_i = \rho_i^2$, and statistical probabilities that reflect computational and other practical difficulties. The recognition of such operators allows the unification of quantum theory and general thermodynamics for all systems (both macroscopic and microscopic), and for all states (both thermodynamic equilibrium and not thermodynamic equilibrium). Moreover, and perhaps more importantly, in contrast to the Schrödinger equation which describes only reversible adiabatic processes that correspond to unitary transformations of projectors, the density operators that are devoid of statistical probabilities provide the basis for the postulation of the complete equation of motion that accounts for all reversible adiabatic processes – both unitary and nonunitary – and all irreversible processes (see Appendix).

APPENDIX A: QUANTUM THEORY

In this Appendix we present a summary of nonrelativistic quantum theory that differs from the presentations in most textbooks on the subject. The key differences are the discoveries that for a broad class of quantum-mechanical problems: (*i*) the probabilities associated with ensembles of measurement results at an instant in time require a mathematical concept delimited by but more general than a wave function or projector; and (*ii*) the evolution in time of the new mathematical concept requires an equation of motion that differs from the correct but incomplete Schrödinger equation.

1.  Definitions, postulates, and theorems at an instant in time

*System*. The term *system* means a set of specified types and amounts of constituents, confined by a nest of internal and external forces. For example, one hydrogen atom consisting of an electron and a proton confined in a one-dimensional square potential well of width equal to $2a$ and height equal to $\infty$. The internal force arises from the Coulomb interaction between the proton and the electron, and depends on the spatial coordinates of both constituents. The external force arises from the gradient of the potential energy, and depends only on the spatial coordinates of either electron, or the proton, or the hydrogen atom as a unit, and not on any coordinates of constituents that are not included in the system, that is, the system is separable from its environment. In addition, in order to be totally independent and fully identifiable, the system must also be *statistically uncorrelated* with its environment.

*System postulate*. To every system there corresponds a complex, separate, complete, inner product space, a Hilbert space $\mathcal{H}$. The Hilbert space of a composite system of two distinguishable and identifiable systems 1 and 2, with associated Hilbert spaces $\mathcal{H}^1$ and $\mathcal{H}^2$, respectively, is the direct product space $\mathcal{H}^1 \otimes \mathcal{H}^2$.

*Homogeneous* or *unambiguous ensemble*. At an instant in time, an ensemble of identical systems is called *homogeneous* or *unambiguous* only if upon subdivision into subensembles in any conceivable way that does not perturb any member, each subensemble yields in every respect measurement results – spectra of values and frequency of occurrence of each value within a spectrum – identical to the corresponding results obtained from the ensemble. For example, the spectrum of energy measurement results and the frequency of occurrence of each energy measurement result obtained from any subensemble are identical to the spectrum of energy measurement results and the frequency of occurrence of each energy measurement result obtained from an independent ensemble that includes all subensembles.

*Preparation*. A *preparation* is a reproducible scheme used to generate one or more homogeneous ensembles for study.

*Property*. The term *property* refers to any attribute of a system that can be quantitatively evaluated at an instant in time by means of measurements and specified procedures. All measurement results and procedures are assumed to be precise, that is, to be both error free, and not to have been affected by the measurement and the measurement procedures. Moreover, they are assumed not to depend on either other systems or other instants in time.

*Observable*. From the definitions just cited, it follows that each property can be observed. Traditionally, however, in quantum theory a property is called an *observable* only if it conforms to the following mathematical representations.

*Correspondence postulate*. Some linear Hermitian operators A, B, … on Hilbert space $\mathcal{H}$, which have complete orthonormal sets of eigenvectors, correspond to observables of a system.

The inclusion of the word some in the correspondence postulate is very important because it indicates that there exist Hermitian operators that do not represent observables, and properties that cannot be represented by Hermitian operators. The first category accounts for Wick *et al*. [24] superselection rules, and the second for compatibility of simultaneous measurements introduced by Park and Margenau [1], and properties, such as temperature, that are not represented by an operator.

*Measurement act*. A *measurement act* is a reproducible scheme of measurements and operations on a member of an ensemble. If the measurement refers to an observable, the result of such an act is a precise, error and perturbation free number associated with the observable.

If a measurement act is applied to each and every member of a homogeneous ensemble, the results conform to the following postulate and theorems.

*Mean-value postulate*. If a measurement act of an observable represented by Hermitian operator A is applied to each and every member of a homogeneous ensemble, there exists a linear functional m(A) of A such that the value of m(A) equals the arithmetic mean of the ensemble of A measurements, that is,

$$m(A) = \langle A \rangle = \sum_i a_i / N \text{ for } N \to \infty \tag{A1}$$

where $a_1$ is the measurement result of the measurement act applied to the ith member of the ensemble, a large number of $a_i$'s have the same numerical value, and both m(A) and $\langle A \rangle$ represent the expectation value of A.

*Mean-value theorem*. For each of the mean-value functionals or expectation values m(A) of a system at an instant in time, there exists the same Hermitian operator ρ such that

$$m(A) = \langle A \rangle = \text{Tr}[\rho A] \tag{A2}$$

The operator ρ is known as the *density operator* or the *density of measurement results of observables*. In view of the definition of a homogeneous ensemble, a pictorial representation of an ensemble that represents $\rho$ is as shown in Figure 2, that is, each member of the ensemble is characterized by the same $\rho$ as any other member.

The concept of the density operator $\rho > \rho^2$ was introduced by von Neumann as a statistical average of projectors. Such a statistical average is called heterogeneous (Figure 1). In contrast, here $\rho$ is restricted to homogeneous ensembles and, therefore, it is exclusively quantum-theoretic, that is, involves no statistical probabilities. The importance of this distinction cannot be overemphasized.

The operator $\rho$ is proven to be Hermitian, positive, unit trace and, in general, not a projector, that is,

$$\rho > 0; \ \text{Tr}\rho = 1; \ \text{and} \ \rho \geq \rho^2 \qquad (A3)$$

The existence of density operators that satisfy all the requirements of quantum theory, and that can be represented only by homogeneous ensembles has been discovered by Hatsopoulos and Gyftopoulos [2], and Jauch [3].

*Probability theorem.* If a measurement act of an observable represented by operator A is applied to each and every member of a homogeneous ensemble characterized by $\rho$, the probability or frequency $W(a_n)$ that the results will yield eigenvalue $a_n$ is given by the relation

$$W(a_n) = \text{Tr}[\rho A_n] \qquad (A4)$$

where $A_n$ is the projection onto the subspace belonging to $a_n$

$$A|\alpha_n^{(i)}\rangle = a_n |\alpha_n^{(i)}\rangle \ \text{for n = 1, 2, ..., and i = 1, 2, ..., g,} \qquad (A5)$$

and g is the degeneracy of $a_n$.

*Measurement result theorem.* The only possible result of a measurement act of the observable represented by A is one of the eigenvalues of A (Eq. A5).

*State*. The term *state* means all that can be said about a system at an instant in time, that is, a set of Hermitian operators A, B, ... that correspond to a set of $n^2 - 1$ independent observables – the values of an independent observable can be varied without affecting the values of other observables – and the relations

$$\langle A \rangle = \text{Tr}[\rho A] = \Sigma_i a_i / N$$

$$\langle B \rangle = \text{Tr}[\rho B] = \Sigma_i b_i / N \tag{A6}$$

$$\vdots \quad \vdots \quad \vdots \quad \vdots \quad \vdots$$

where n is the dimensionality of the Hilbert space.

In Eqs. (A6), either the density operator $\rho$ is specified *a priori* and the values of the observables are calculated, or the values of the independent observables $\Sigma_i a_i / N$, $\Sigma_i b_i / N$, …, are either specified or, in principle, experimentally established, and a unique density operator is calculated. The mappings from $\rho$ to expectation values, and from expectation values to $\rho$ are unique because Eqs. (A6) are linear from expectation values to $\rho$ and from $\rho$ to expectation values.

It is noteworthy, that no quantum-theoretic requirement exists which excludes the possibility that the mapping from expectation values to $\rho$ must yield a projector $\rho_i = \rho_i^2$ rather than a density operator $\rho > \rho^2$. In fact, the existence of density operators that are not derived as a mixture of quantum probabilities and statistical probabilities provides the means for the unification of quantum theory and thermodynamics without resorting to statistical considerations [2, 25-27].

It is also noteworthy that only the first power of an operator X and its eigenvalues $x_i$ are included in Eqs. (A6). For example, only the Hamiltonian operator H and its eigenvalues $\varepsilon_1$, $\varepsilon_2$, …, appear once in Eqs. (A6). Operators $H^m$ and their eigenvalues $\varepsilon_1^m$, $\varepsilon_2^m$, … for $m > 1$ are excluded. The reason for this important restriction is that information about the sign of eigenvalues of any X is lost.

2. Evolution of the density operator in time

*Dynamical postulate.* Hatsopoulos and Gyftopoulos [25] postulate that unitary transformations of $\rho$ in time obey the relation

$$\frac{d\rho}{dt} = -\frac{i}{\hbar}[H\rho - \rho H] \tag{A7}$$

where H is the Hamiltonian operator of the system. The unitary transformation of $\rho$ satisfies the equation

$$\rho(t) = U(t, t_0) \, \rho(t_0) U^+(t, t_0) \tag{A8}$$

where $U^+$ is the Hermitian conjugate of U and, if H is independent of $t$,

$$U(t,\ t_0) = \exp\left[-(i/\hbar)(t-t_0)H\right] \tag{A9}$$

and, if H is explicitly dependent on $t$,

$$\frac{dU(t,\ t_0)}{dt} = -(i/\hbar)H(t)U(t,\ t_0) \tag{A10}$$

Though Eq. (A7) is well known in the literature as the von Neumann equation, here it must be postulated for the following reason. In statistical quantum mechanics [28] the equation is derived as a statistical average of Schrödinger equations, each of which describes the evolution in time of a projector $\rho_i$ in the statistical mixture represented by $\rho$, and each of which is multiplied by a time independent statistical probability $\alpha_i$. But here, $\rho$ is not a mixture of projectors and, therefore, cannot be derived as a statistical average of projectors. It is noteworthy that the dynamical postulate is limited or incomplete because all unitary evolutions of $\rho$ in time correspond to reversible adiabatic processes, but not all reversible adiabatic processes correspond to unitary evolutions of $\rho$ in time [27], and not all processes are reversible.

An equation that describes both all reversible processes and irreversible processes has been conceived by Beretta *et al*. [29, 30]. It is not discussed here because it is outside the scope of this article. Nevertheless, it is noteworthy that the Beretta equation is shown to satisfy all the requirements for it to be a bona fide equation of motion of a nonstatistical unified theory of quantum mechanics and thermodynamics [31].

**List of Figures**

Figure 1. Pictorial representation of a heterogeneous ensemble. Each of the subensembles for $\rho_1$ and $\rho_2$ can represent either a projector ($\rho_i = \rho_i^2$) or a density operator ($\rho_i > \rho_i^2$).

Figure 2. Pictorial representation of a homogeneous ensemble. Each of the members of the ensemble is characterized by the same density operator $\rho \geq \rho^2$. It is clear that any conceivable subensemble of a homogeneous ensemble is characterized by the same $\rho$ as the ensemble.